\documentclass[epj]{svjour}
\usepackage{latexsym}
\usepackage{graphics}
\usepackage{amsmath}
\usepackage{color}
\usepackage{graphicx}

\begin{document}

\title{Molecular dynamics study of the stability of a carbon nanotube atop a catalytic nanoparticle}

\author{
Alexey V. Verkhovtsev\inst{1}
%\thanks{verkhovtsev@fias.uni-frankfurt.de,
\thanks{\email{verkhovtsev@fias.uni-frankfurt.de},
On leave from A.F. Ioffe Physical-Technical Institute, St. Petersburg, Russia}
\and
Stefan Schramm\inst{1,2}
\and
Andrey V. Solov'yov\inst{3,2}
\thanks{On leave from A.F. Ioffe Physical-Technical Institute, St. Petersburg, Russia}
}

\institute{
Frankfurt Institute for Advanced Studies, Ruth-Moufang-Str. 1, 60438 Frankfurt am Main, Germany
\and
Department of Physics, Goethe University Frankfurt, Max-von-Laue-Str. 1, 60438 Frankfurt am Main, Germany
\and
MBN Research Center, Frankfurter Innovationszentrum, Altenh\"oferallee 3, 60438 Frankfurt am Main, Germany
}

\date{\today}

\abstract{
The stability of a single-walled carbon nanotube placed on top of a catalytic nickel
nanoparticle is investigated by means of molecular dynamics simulations.
As a case study, we consider the $(12,0)$ nanotube consisting of 720 carbon atoms and
the icosahedral Ni$_{309}$ cluster.
An explicit set of constant-temperature simulations is performed in order to cover a broad
temperature range from 400 to 1200~K, at which a successful growth of carbon nanotubes has
been achieved experimentally by means of chemical vapor deposition.
The stability of the system depending on parameters of the involved interatomic
interactions is analyzed.
It is demonstrated that different scenarios of the nanotube dynamics atop the nanoparticle
are possible depending on the parameters of the Ni--C potential.
When the interaction is weak the nanotube is stable and resembles its highly symmetric
structure, while an increase of the interaction energy leads to the abrupt collapse of the
nanotube in the initial stage of simulation.
In order to validate the parameters of the Ni--C interaction utilized in the simulations,
DFT calculations of the potential energy surface for carbon-nickel compounds are performed.
The calculated dissociation energy of the Ni--C bond is in good agreement with the values,
which correspond to the case of a stable and not deformed nanotube simulated within the
MD approach.
}

\authorrunning{Verkhovtsev, Schramm, Solov'yov}
\titlerunning{MD study of the stability of a carbon nanotube atop a catalytic nanoparticle}

\maketitle

%%%%%%%%%%%%%%%%%%%%%%%%%%%%%%%%%%%%%%%%%%%%%%%%%%%%%%%
\section{Introduction}

During past decades, mass production synthesis of carbon nanotubes \cite{Iijima_1991_Nature.354.56}
has been achieved by various methods such as arc discharge \cite{Journet_1997_Nature.388.756},
laser ablation \cite{Smalley_1996_Science.273.483}, and pyrolysis \cite{Terrones_1997_Nature.388.52}.
In these methods, the growth of nanotubes is initiated by condensation of a hot carbon-rich gas.
These approaches does not allow for an easy control of the diameter, length and chirality
of nanotubes that strongly affect physical and mechanical properties of the systems
\cite{Choi_2000_ApplPhysLett.76.2367}.
Thus, different variations of the chemical vapor deposition (CVD) method have been utilized
in order to synthesize nanotubes of higher purity and with more selective growth parameters
\cite{Choi_2000_ApplPhysLett.76.2367,Li_2008_MaterLett.62.1472,Yuan_2008_NanoLett.8.2576,Kang_2009_JMaterSci.44.2471,Diao_2010_AdvMater.22.1430}.
In this method, the nanotube growth is determined by the presence of a catalytic
nanoparticle (usually made of transition elements like Ni, Co and Fe), which causes the
separation of carbon present in the precursor gas (e.g., CO, C$_2$H$_2$, or CH$_4$) and
then assists as seed-site for the growth \cite{Martinez-Limia_2007_JMolModel.13.595}.

It has been found that the structure of carbon nanotubes is dependent on synthesis parameters
such as the reaction temperature \cite{Li_2008_MaterLett.62.1472}, the composition and size
of a catalytic particle
\cite{Choi_2000_ApplPhysLett.76.2367,Yuan_2008_NanoLett.8.2576,Harris_2007_Carbon.45.229,Kang_2008_MaterSciEngA.475.136},
the reaction gas \cite{Harris_2007_Carbon.45.229,SaitoBook}, etc.
In spite of an intense study of the growth mechanism of carbon nanotubes, its details are
not yet fully understood, so the current synthesis methods do not allow for a full control
of the growth \cite{Harutyunyan_2005_ApplPhysLett.87.051919}.
In particular, there has been much discussion during past years on the phase of catalysts
in the course of the nanotube synthesis
\cite{Harutyunyan_2005_ApplPhysLett.87.051919,Kanzow_1999_PhysRevB.60.11180,Gavillet_2001_PhysRevLett.87.275504,Ding_2004_JChemPhys.121.2775}.
Typically, the process of the nanotube growth by CVD methods is conducted at elevated
temperatures approximately ranging from 900 to 1500~K
\cite{Harutyunyan_2005_ApplPhysLett.87.051919,Cui_2000_JApplPhys.88.6072,Jeong_2003_JpnJApplPhys.42.L1340}.
Although these values are significantly lower than the melting temperature of bulk metals
that typically comprise the catalyst, the nanoparticles can be in the molten state due to
%the well-established linear decrease of the melting temperature of small nanoparticles
%with their inverse radius
the well-established dependence of the melting temperature of small nanoparticles on their
radius, $T_m = T_m^{\rm bulk}\left( 1 - \alpha/R \right)$,
where $T_m^{\rm bulk}$ is the bulk melting temperature,
$R$ is the radius of a spherical nanoparticle and $\alpha$ is the factor of
proportionality
\cite{Pawlow_1909_ZPhysChem.65,Qi_2001_JChemPhys.115.385,Lyalin_2009_PhysRevB.79.165403,Yakubovich_2013_PhysRevB.88.035438}.
In Ref. \cite{Harutyunyan_2005_ApplPhysLett.87.051919} it was found that the liquid phase of an
iron catalytic nanoparticle is favored for the growth of nanotubes, while the solidification
of the catalyst nearly terminates the growth.
On the other hand, a number of successful attempts to lower the nanotube synthesis temperature
by the plasma-enhanced CVD method have been reported
\cite{Hofmann_2003_ApplPhysLett.83.135,Shang_2010_Nanotech.21.505604,He_2011_NanoRes.4.334,Halonen_2011_PhysStatSolb.248.2500}.
In particular, this method has allowed one to grow vertically aligned nanotubes at temperatures
as low as 400~K with nickel as a catalyst \cite{Hofmann_2003_ApplPhysLett.83.135}.
In the case of such a low-temperature growth, the catalytic nanoparticle certainly remains in
the solid state.

%As briefly described above, experimental studies provide some details of the carbon nanotube
%growth mechanism.
%However, experimental observations cannot explain the atomistic details of the process such
%as mechanisms of graphitic cap nucleation on the cluster surface, how the open end of the
%nanotube is maintained during the growth, how defects that may form in the nanotube are healed,
%and what determines the diameter and chirality of the nanotube.

In spite of intensive research and a huge amount of collected experimental information,
the physical mechanisms leading to the catalytically-assisted carbon nanotube growth remain
a highly debated issue.
A limited understanding of the detailed growth mechanism hinders further progress in the carbon
nanotube production such as the selective growth of nanotubes having a specific diameter and/or
chirality \cite{Diao_2010_AdvMater.22.1430,Harris_2007_Carbon.45.229}.

Numerical studies based on molecular dynamics (MD) simulations have provided deeper understanding
of the catalyzed nanotube growth and its initial stage, in particular, on the atomistic level
\cite{Ding_2004_JChemPhys.121.2775,Irle_2009_NanoRes.2.755,Ohta_2009_Carbon.47.1270,Shibuta_2003_ChemPhysLett.382.381,Zhao_2005_Nanotechnology.16.S575}.
The performed simulations have illustrated the diffusion of carbon atoms into the catalytic
nanoparticle that is followed by the cap formation process when the concentration of carbon
atoms in the catalyst exceeds some critical value.
Classical MD simulations, based on the so-called reactive empirical bond-order (REBO)
potential \cite{Tersoff88,Brenner90}, of the nanotube growth on a metal catalyst have
shown that the following three stages of the process can be distinguished
\cite{Ding_2004_JChemPhys.121.2775,Shibuta_2003_ChemPhysLett.382.381}.
At first, carbon atoms dissolve in the metal nanoparticle, then a small graphitic island
starts to nucleate on the cluster surface.
Finally, the metal nanoparticle becomes covered by carbon atoms and, depending on temperature,
a graphite sheet or a carbon nanotube is formed.
%The diameter of the nanotube in such simulation strongly depend on the size of the catalytic
%particles, it is in agreement with the experimental observations.
However, the resulting structure of modeled nanotubes contained a significant number of
defects, such as four-, five-, seven- or eight-membered rings, that made the prediction of the
nanotube chirality hardly possible.

%One of the major tasks of the nanotube production is related to the control of structural properties
%of the nanotubes such as the diameter and chirality \cite{Wang05, Smalley06, Ohta08, Irle09}.
In the meantime, Smalley and co-workers have reported \cite{Smalley_2006_JACS.128.15824} on the
successful experimental attachment of a short single-walled nanotube to iron nanoparticles and
showed that continued growth of the nanotube can be achieved.
During the last years, several theoretical studies of the continued growth of a single-walled
carbon nanotube on small iron clusters were performed using quantum MD simulations
\cite{Irle_2009_NanoRes.2.755,Ohta_2008_ACSNano.2.1437}.
In these studies, it was shown that due to the random nature of new polygon formation, the tube
sidewall growth is observed as an irregular process without clear nanotube chirality.

In Ref. \cite{Solovyov_2008_PhysRevE.78.051601} the so-called liquid surface model for calculating
the energy of single-walled carbon nanotubes with arbitrary chirality was introduced.
The model allowed one to predict the energy of a nanotube once its chirality and the total number
of atoms are known.
The suggested model gave an insight in the energetics and stability of nanotubes of different
chirality and was considered as an important tool for the understanding of the nanotube growth.
%The model also gave estimates of the influence of the catalytic nanoparticle, atop which nanotubes
%grow, on the nanotube stability.

In spite of an intensive experimental and theoretical research carried out within the past
few decades, details of the carbon nanotube stabilization mechanism on metal nanoparticles
are not yet completely understood, although they can strongly affect the continued growth scenario.
In the present study, we investigate the process of the carbon nanotube stabilization on
a catalytic nickel nanoparticle by means of classical MD simulations.
In particular, we analyze how the reaction temperature affects the nanotube stability.
For this purpose, we have performed a set of constant-temperature simulations, which cover
the whole range of temperatures currently achieved in CVD and PECVD methods.
A very important problem of the numerical method, based on MD simulations, is a proper choice
of interatomic potentials and their parameters.
Thus, we consider a broad range of parameters for the nanotube-catalyst interaction and demonstrate
that, depending on the parameters of a Ni--C interatomic potential, different scenarios of the
nanotube dynamics atop the nanoparticle are possible.
In order to validate the parameters of the Ni--C interaction, utilized in the simulations,
we perform a set of {\it ab initio} calculations of the potential energy surface for two
Ni--C-based compounds,
namely for a Ni atom linked to a benzene molecule and to a small carbon chain.
Considering these case studies, we have accounted for different types of Ni--C interaction,
that exist when nickel atoms interact either with the nanotube sidewall or with its open end.
The resulting dissociation energy of the Ni--C bond is in good agreement with the values,
which are utilized in MD simulations describing the stable system.

%%%%%%%%%%%%%%%%%%%%%%%%%%%%%%%%%%%%%%%%%%%%%%%%%%%%%%%
\section{Theoretical framework}
\subsection{Description of the system}

In this study, we analyze the stability of a single-walled carbon nanotube placed atop a
nickel nanoparticle by means of MD simulations.
As a case study, we consider a 6.24~nm-long uncapped nanotube of the $(12,0)$ chirality
consisting of 720 carbon atoms.
The diameter of the nanotube is equal to 0.94~nm.
In its initial configuration, the nanotube is aligned along the $z$ axis, as illustrated
in Figure~\ref{Fig_overview}(a).
The geometrical structure of the nanotube was obtained using the Nanotube Builder tool of
the VMD program \cite{VMD_reference}, which was also used for the visualization of the results.
The nanotube is positioned over the top face of the icosahedral Ni$_{309}$ cluster, which has
a radius of approximately 1.8~nm.
The initial structure of the system under study is presented in Figure~\ref{Fig_overview}(a).

In order to investigate the nanotube dynamics on top of the catalytic nickel nanoparticle,
we have conducted an explicit set of constant-temperature MD simulations.
They were performed using MBN Explorer \cite{MBN_Explorer1,MBN_Explorer2}, a universal software
package for multiscale simulation of complex molecular structure and dynamics.
During past years it has been utilized for structure optimization
\cite{Solovyov_2003_PhysRevLett.90.053401,Geng_2010_PhysRevB.81.214114},
simulation of dynamics
\cite{Yakubovich_2013_PhysRevB.88.035438,Verkhovtsev_2013_ComputMaterSci.76.20,Yakubovich_2013_ComputMaterSci.76.60,Sushko_2013_JCompPhys.252.404}
and growth processes \cite{Dick_2011_PhysRevB.84.115408,Solovyov_2013_PhysStatSolB,Panshenskov_3d-KMC}
in various molecular and bio/nanosystems.
Integration of equations of motion was done using the Verlet leap frog algorithm with a
time step of 1~fs and a total simulation time of 5~ns.
The total energy of the system and coordinates of all atoms were recorded each 1~ps of
the simulation.
The temperature control was achieved by means of the Langevin thermostat with a damping
constant $\gamma = 0.1$~ps$^{-1}$.
In this case, the dynamics of atoms in the system is described by Langevin equations of motion:
\begin{equation}
m_i {\bf a}_i(t) =
{\bf F}_i(t) - \frac{1}{\tau_d} m_i{\bf v}_i(t) + \sqrt{\frac{2 k_B T_0 m_i}{\tau_d} } \, {\bf R}_i(t) \ ,
\label{Langevin}
\end{equation}

\noindent
where ${\bf F}_i$ is the physical force acting on the atom,
$k_B T$ denotes the thermal energy in the system,
$\tau_d \equiv 1/\gamma$ is the characteristic viscous damping time, and
${\bf R}_i(t)$ represents
random forces, which act on the particle as a result of solvent interaction.
%a delta-correlated stationary Gaussian process with zero-mean, satisfying
%
%\begin{equation}
%\langle {\bf R}_i(t) \rangle = 0
%\end{equation}
%\begin{equation}
%\langle {\bf R}_i(t) \, {\bf R}_i(t^{\prime}) \rangle = \delta(t - t^{\prime}) \ ,
%\end{equation}
%
%\noindent
%where $\langle \dots \rangle$ denotes time-averaging.
The Langevin equation of motion gives a physically correct description of a many-particle
system interacting with a heat bath maintained at a constant temperature $T_0$ \cite{MBN_Explorer1}.

Calculations performed within the classical framework were carried out using empirical
interatomic potentials.
The total potential energy of the system is defined as follows:
\begin{equation}
U = U^{\rm Ni-Ni} + U^{\rm C-C}+U^{\rm Ni-C} \ ,
\label{eq:totalPotentialEnergy}
\end{equation}

\noindent
where the terms $U^{\rm Ni-Ni}$, $U^{\rm C-C}$ and  $U^{\rm Ni-C}$ refer to the nickel-nickel,
carbon-carbon and nickel-carbon interaction, respectively.
A detailed description of the potentials used in the calculations is given in the following
subsections.

%%%%%%%%%%%%%%%%%%%%
\subsection{Nickel-nickel interaction}

The interaction between nickel atoms is described in the present study using the many-body
Finnis-Sinclair-type potential \cite{Finnis-Sinclair}.
Recently, this interatomic potential has been successfully utilized for studying the
diffusion process in both ideal and nanostructured titanium and nickel-titanium crystalline
samples \cite{Yakubovich_2013_ComputMaterSci.76.60,Sushko_2014_JPhysChemA2}, as well as
for analysis of mechanical properties of these materials by means of MD simulations of
nanoindentation
\cite{Verkhovtsev_2013_ComputMaterSci.76.20,Verkhovtsev_2013_ComputTheorChem.1021.101,Sushko_2014_JPhysChemA}.

%Besides the Finnis-Sinclair (FS) method, another most widely used potential format for
%metallic systems is given by the embedded atom method (EAM)
%\cite{Daw_1983_PhysRevLett.50.1285, Daw_1984_PhysRevB.29.6443}.
%Although initially derived from different physical approaches (density-functional theory
%for EAM and second-moment tight binding for FS), the two potential forms are similar
%\cite{Mishin_2010_ActaMater.58.1117}.

The general structure of many-body potentials
\cite{Finnis-Sinclair,Gupta,Sutton_Chen,Daw_1993_MaterSciRep.9.251,Rafii-Tabar_potentials,TB-SMA}
contains an attractive density-dependent many-body term and a repulsive part for small distances
that results from the repulsion between core electrons of neighboring atoms.

In the Finnis-Sinclair representation, the total energy of an $N$-atom system is written as:
\begin{equation}
U_{\rm FS} = \frac12 \sum_{i=1}^N\sum_{j \ne i} V(r_{ij}) - c \sum_i \sqrt{ \rho_i } \ ,
\label{FS}
\end{equation}
where
\begin{equation}
\rho_i = \sum_{j \ne i} \phi(r_{ij}) \ .
\label{FS_2}
\end{equation}

\noindent
Here the function $V(r_{ij})$ is a pairwise repulsive interaction between atoms $i$ and $j$
separated by a distance $r_{ij}$, the function $\phi(r_{ij})$ describes an attractive
pair potential, and $c$ is a positive constant.
The second term in Eq.~(\ref{FS}) represents the attractive many-body contribution to
the total energy of the system.
The square root form of this term is chosen in the FS approach in order to mimic the
result of tight-binding theory, in which $\phi(r)$ is interpreted as a sum of squares
of overlap integrals \cite{Finnis-Sinclair}.
According to this approach (see, e.g., \cite{Rafii-Tabar_potentials,TB-SMA,TB-SMA_2}
and references therein), the energy of the $d$ electron band in metals
\begin{equation}
E_{\rm band} = \sum_i E_i = 2\sum_i \int\limits_{-\infty}^{E_F} E \, n_i(E) \, {\rm d} E
\label{TB_Eband}
\end{equation}

\noindent
is proportional to the square root of the second moment of the density of states.
Here $E_i$ is the contribution to the total electronic band energy from
an individual atom $i$, $n_i(E)$ the density of states projected on site $i$,
$E_F$ the Fermi level energy, and prefactor 2 in Eq.~(\ref{TB_Eband}) arises due to
spin degeneracy \cite{Rafii-Tabar_potentials}.
The second moment $\mu_2^i$ of the density of states, defined as
\begin{equation}
\mu_2^i =  \int\limits_{-\infty}^{+\infty} E^2 \, n_i(E) \, {\rm d} E \ ,
\end{equation}

\noindent provides a measure of the squared band width and allows one to derive
an approximate expression for the band energy in terms of $\mu_2^i$.
This function can also be expressed as
\begin{equation}
\mu_2^i = \sum_j h_{ij}^2 \ ,
\end{equation}

\noindent where
\begin{equation}
h_{ij} = \langle \chi_i \, | \, H \, | \, \chi_j \rangle  \ ,
\end{equation}

\noindent $\chi_i$ is the localized orbital centered on atom $i$, and
$H$ the single-electron Hamiltonian.
The electron band energy $E_i$ is then expressed as
\begin{equation}
E_i \equiv E_i(\mu_2^i) =  -A\sqrt{(\mu_2^i)} = -A\sqrt{\sum_j h_{ij}^2} \ ,
\end{equation}

\noindent where $A$ is a positive constant that depends on the chosen density of
states shape (see Ref. \cite{Rafii-Tabar_potentials} and references therein).
The square root form of this term is chosen since $E_i$ has units of energy and
$\mu_i^2$ has units of energy squared.

The functions $\phi(r_{ij})$ in Eq.~(\ref{FS_2}) can thus be interpreted as the sum of
squares of overlap integrals \cite{Rafii-Tabar_potentials,Li_2007_JPhysCondMatter.19.086228}.
The function $\rho_i$ in the FS approach can be interpreted as a measure of the local
density of atomic sites \cite{Finnis-Sinclair}.
Note that the Finnis-Sinclair potential, Eq.~(\ref{FS}), is similar in form to the embedded-atom
model (EAM) potential \cite{Daw_1993_MaterSciRep.9.251}, although the interpretation of the
function $\rho_i$ is different in the two cases.
In the EAM approach, $\rho_i$ stands for the local electronic charge density at site $i$
constructed by a rigid superposition of atomic charge densities $\phi(r_{ij})$.
In other words, $\rho_i$ is the host electron density induced at site $i$ by all other atoms.
In this case, the energy of an atom at site $i$ is assumed to be identical to
its energy within a uniform electron gas of that density
\cite{Daw_1983_PhysRevLett.50.1285,Daw_1984_PhysRevB.29.6443}.

Similar to the original second-moment approximation of the tight-binding (TB-SMA) scheme,
the functions $V(r_{ij})$ and $\phi(r_{ij})$ in Eq.~(\ref{FS}) and (\ref{FS_2}) are
introduced in exponential forms \cite{TB-SMA,TB-SMA_2}.
As indicated in Ref.~\cite{TB-SMA}, the standard dependence of the band energy on
radial interatomic distance between atoms $i$ and $j$ should rather be proportional
to $r^{-4}$ or $r^{-5}$, although an exponential form of this dependence better
accounts for atomic relaxation near impurities and surfaces \cite{Tomanek_1985_PhysRevB.32.5051}.

Finally, the total potential energy, $U \equiv U^{\rm Ni-Ni}$, of a system of $N$ nickel atoms,
located at positions ${\bf r}_i$, in the Finnis-Sinclair representation reads as:
%
%\begin{eqnarray}
\begin{equation}
%U^{\rm Ni-Ni} =
U =
\sum_{i=1}^N \left[ \sum_{\substack{j=1 \\ (i \ne j)}}^N A\,
{\rm e}^{-p \left( \frac{r_{ij}}{d} - 1 \right)} \right.
%\nonumber \\
- \left. \sqrt{ \sum_{\substack{j=1 \\ (i \ne j)}}^N \xi^2\,
{\rm e}^{-2q \left( \frac{r_{ij}}{d} - 1 \right)}  } \, \right] \ ,
\label{Finnis_Sinclair_pot}
%\end{eqnarray}
\end{equation}

\noindent where $r_{ij}$ is the distance between atoms $i$ and $j$, and
$d$, $p$, $q$, $A$ and $\xi$ are adjustable parameters of the potential.
The parameter $d$ is the first-neighbor distance,
$\xi$ results from an effective hopping integral, $q$ describes its dependence
on the relative interatomic distance, and the parameter $p$ is related to the
compressibility of the bulk metal \cite{TB-SMA}.

Note that the Finnis-Sinclair-type potential, as implemented in MBN Explorer
\cite{MBN_Explorer1,MBN_Explorer2}, can be applied not only for monoatomic systems,
but also for bimetallic compounds
\cite{Verkhovtsev_2013_ComputTheorChem.1021.101,Sushko_2014_JPhysChemA}.
In the latter case, the aforementioned parameters, $d \equiv d_{\alpha\beta}$, etc.,
depend on the type of an atom, $\alpha / \beta$, chosen within the summation.
When $\alpha = \beta$, as in the present study where only the Ni--Ni interaction is
described, such a type of the potential is also referred to in the literature as the
Gupta potential \cite{Gupta}.

In order to describe the interaction between nickel atoms, we used the parametrization
introduced in Ref. \cite{Lai_2000_JPhysCondMatter.12.L53} that reproduces main
mechanical and structural properties
%, such as the cohesive energy, lattice parameters and elastic constants,
of bulk nickel crystal at zero temperature.
The parameters provided for nickel have the following values:
$d = 2.49$~\AA, $A = 0.104$~eV, $p = 11.198$, $\xi = 1.591$~eV, and $q = 2.413$
\cite{Lai_2000_JPhysCondMatter.12.L53}.

Since most of the many-body potentials approach zero at large distances,
a cutoff radius $r_c$ is frequently introduced to reduce the computation time.
In this case, the interatomic potentials and, subsequently, the forces are neglected
for atoms positioned at distances larger than $r_c$ from each other.
The utilized parameter set \cite{Lai_2000_JPhysCondMatter.12.L53} was constructed
with a fixed cutoff radius of 4.2~\AA.
In order to avoid the effect of non-continuity of the potential due to its non-zero
value at the cutoff radius, we have implemented a polynomial switching from the
original potential value at the cutoff radius of 4.2~\AA, to zero value at the
extended cutoff radius of 5.5~\AA.
Coefficients of the splines were determined to correspond to the value and the first
derivative of the potential at the initial cutoff and to be equal to zero at 5.5~\AA \ \cite{Verkhovtsev_2013_ComputTheorChem.1021.101}.

%%%%%%%%%%%%%%%%%%%%
\subsection{Carbon-carbon interaction}

In order to describe the carbon-carbon interaction, we employed the Brenner empirical
potential \cite{Brenner90}, which was developed for studying carbon-based systems with
different types of covalent bonds.
It is a REBO-type potential, which is able to account for the bond breaking/formation
using a distance-dependent many-body order term and correctly dissociating diatomic
potentials \cite{Irle_2009_NanoRes.2.755}.

For every atom in the system, this many-body potential depends on the nearest neighbors
of this atom.
The total energy, $U^{\rm C-C}$, of the system of $N$ carbon atoms interacting via the
Brenner potential is expressed as a sum of bonding energies between all atoms:
\begin{eqnarray}
U^{\rm C-C} &=& \frac12 \sum_{i=1}^N \sum_{\substack{j=1 \\ (i \ne j)}}^N  U_{ij}
\\
&=&
\frac12 \sum_{i}\sum_{j\neq i} f_{\rm cut}(r_{ij})
\left[ U^{\rm (R)}(r_{ij})-B_{ij} U^{\rm (A)}(r_{ij}) \right] \nonumber \ ,
\label{eq:BondOrder}
\end{eqnarray}

\noindent
where $f_{\rm cut}(r_{ij})$ is the cutoff function, which limits the interaction of an atom to
its nearest neighbors:
\begin{equation}
f_{\rm cut}(r_{ij}) =
\left\{
\begin{array}{ll}
  1 & , \ r_{ij} \leq R_1 \\
  \displaystyle{ \frac12 }
  \left[ 1 + \cos{ \displaystyle{ \left( \frac{r_{ij} - R_1}{R_2 - R_1} \pi \right)} }  \right]
  & , \ R_1 < r_{ij} \leq R_2 \\
  0 & , \ r_{ij} > R_2
\end{array}
\right.
\label{eq:BOcutoff}
\end{equation}

\noindent
with $R_1$ and $R_2$ being the parameters, which determine the range of the potential,
and $r_{ij}$ the distance between atoms $i$ and $j$.
The functions $U^{\rm (R)}(r_{ij})$ and $U^{\rm (A)}(r_{ij})$ are the repulsive and attractive
terms of the potential, respectively.
The Brenner potential implies the following Morse-type exponential parametrization for
these functions:
\begin{eqnarray}
U^{\rm (R)}(r_{ij})&=&\frac{D_{e}}{S-1} \exp \left(- \beta \sqrt{2S} \left(r_{ij}-R_{e} \right)\right) \ ,
\nonumber \\
U^{\rm (A)}(r_{ij})&=& \frac{D_{e}S}{S-1} \exp \left(- \beta \sqrt{2/S} \left(r_{ij}-R_{e} \right)\right) \ ,
\label{eq:BrennerRep+Attr}
\end{eqnarray}

\noindent
where parameters $D_e$, $S$, $\beta$, and $R_e$ are determined from the known physical
properties of carbon, graphite and diamond \cite{Brenner90}.
If $S=2$, then the pair terms (\ref{eq:BrennerRep+Attr})
reduce to the usual Morse potential.
The well depth $D_e$, equilibrium distance $R_e$ and $\beta$, which defines the well width,
are equal to the usual Morse parameters independent of the value of $S$ \cite{Brenner90}.

The factor $B_{ij}$ in Eq.~(\ref{eq:BondOrder}) is the empirical bond-order function,
which is defined as follows:
\begin{equation}
B_{ij} = \left[ 1+ \sum_{k\ne i,j} f_{\rm cut}(r_{ik})G(\theta_{ijk}) \right]^{-\delta} =
\left[ 1+ \zeta_{ij}\right]^{-\delta}
\label{eq:BOfactorb}
\end{equation}

\noindent
Here $\delta$ is the parameter, which may depend on the particular system,
%$f_{C}(r_{ij})$ is the cutoff function introduced in Eq.~(\ref{eq:BOcutoff}),
and the function $G(\theta_{ijk})$ is defined as:
\begin{equation}
G(\theta_{ijk}) =
a \left[ 1+ \frac{c^{2}}{d^{2}}-\frac{c^{2}}{d^{2} + (1 + \cos\theta_{ijk} )^2} \right] \ ,
\label{eq:BOtheta}
\end{equation}

\noindent
where $\theta_{ijk}$ is the angle between bonds formed by pairs of atoms $(i,j)$ and $(j,k)$, so that
\begin{equation}
\cos\theta_{ijk} = \frac{{\bf r}_{ij}\cdot {\bf r}_{jk}}{r_{ij} \, r_{jk}} \ .
\end{equation}

The parameters \cite{Brenner90} of the potential are summarized in Table~\ref{tab:tab1}.
\begin{table}[h]
\centering
\caption{
Parameters of the Brenner potential \cite{Brenner90} utilized for calculation
of the carbon-carbon interaction.}
\begin{tabular}{ll}
\hline
  $D_{e}$ (eV) \ \ \ & 6.325    \\
  $S$                & 1.29     \\
  $\beta$ (1/\AA)    & 1.5      \\
  $R_{e}$ (\AA)      & 1.315    \\
  $R_1$ (\AA)        & 1.7      \\
  $R_2$ (\AA)        & 2.0      \\
  $\delta$           & 0.80469  \\
  $a$                & 0.011304 \\
  $c$                & 19       \\
  $d$                & 2.5      \\
\hline
\end{tabular}
\label{tab:tab1}
\end{table}

%%%%%%%%%%%%%%%%%%%%
\subsection{Nickel--Carbon interaction}

The most crucial issue in the description of stability of a nanotube and its interaction
with a catalytic nanoparticle is a reliable choice of a potential for the metal--carbon
interaction.
The determination of a potential for the transiti- on-metal--carbon (Ni--C, in particular)
interaction is a challenging and nontrivial task.
Nevertheless, a number of various pair and many-body potentials for the description
of such an interaction have been developed so far.

\begin{figure*}
\centering
\resizebox{1.7\columnwidth}{!}{\includegraphics{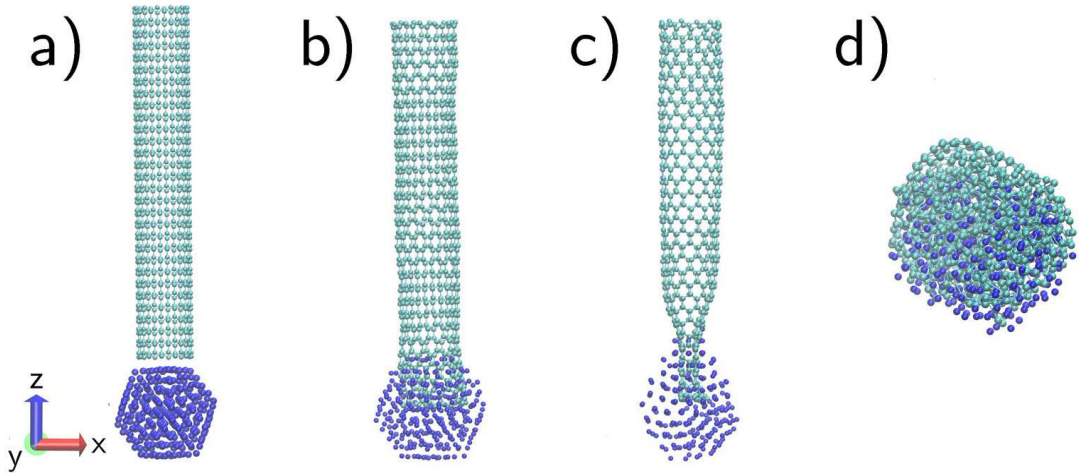}}
\caption{Evolution of the nanotube structure on top of the Ni$_{309}$ nanoparticle at $T = 400$~K.
Panel (a) shows the initial structure of the system.
Panels (b), (c) and (d) illustrate snapshots of the system after 5~ns of simulation at different
values of the binding energy between Ni and C atoms:
(b) $D_e^{\rm Ni-C} = 0.2$~eV: the nanotube placed atop the nanoparticle is stable and resembles
its original structure;
(c) $D_e^{\rm Ni-C} = 0.8$~eV: the nanotube is stable but the structure of both the nanotube and
the nanoparticle is strongly deformed;
(d) $D_e^{\rm Ni-C} = 2.4$~eV: the nanotube is unstable and collapses just after the simulation starts.}
\label{Fig_overview}
\end{figure*}

In Refs. \cite{Yamaguchi_1999_EPJD.9.385,Shibuta_2007_ComputMaterSci.39.842}, a many-body
potential for the transiti- on-metal--carbon interaction was developed on the basis of
density-functional theory (DFT) calculations of small NiC$_n$ ($n=1-3$) clusters.
It was noted that the developed potential was a rough estimation, since the introduced
parametrization was based on the extrapolation of DFT-based results obtained in the case
of small atomic clusters only.
Nevertheless, this potential was utilized for a molecular dynamics study of formation
of metallofullerenes \cite{Yamaguchi_1999_EPJD.9.385} and single-walled carbon nanotubes
\cite{Shibuta_2007_ComputMaterSci.39.842}.
In Ref. \cite{Martinez-Limia_2007_JMolModel.13.595}, a similar approach, based on the DFT
optimization of small Ni--hydrocarbon systems, was utilized for construction of a more
elaborated REBO-type force field, which was applied for simulating the catalyzed growth
of single-walled carbon nanotubes on nickel clusters consisting of several tens of atoms.

Alongside with the bond-order many-body potentials, less sophisticated pairwise potentials
have been also utilized for studying dynamic processes with nickel-carbon systems.
For instance, the pairwise Morse potential was used
in Ref. \cite{Lyalin_2009_PhysRevB.79.165403} to investigate how thermodynamic properties
of a nickel cluster change by the addition of a carbon impurity.
The pairwise Morse potential was chosen for the description of the Ni--C interaction because
of its simplicity. It allowed to study the influence of the parameters on
thermodynamic properties of the C-doped Ni$_{147}$ cluster, maintaining a clear physical
picture of the process occurring in the system \cite{Lyalin_2009_PhysRevB.79.165403}.
For this purpose, the Ni--C interatomic potential, obtained from the results of earlier
DFT calculations of small NiC$_n$ ($n=1-3$) clusters \cite{Yamaguchi_1999_EPJD.9.385},
was fitted by the Morse potential:
\begin{equation}
%U^{\rm Ni-C}(r) = \epsilon_{M}\left(\left(1-e^{\rho(1-r/r_{0}}\right)\right)^{2}-1) \ .
U^{\rm Ni-C}(r) = D_e \left[ \left(1 - e^{-\beta(r - R_e)}\right)^{2} - 1 \right] \ .
\label{eq:eq4}
\end{equation}

\noindent
%and the following parameters were determined:
%$\epsilon_{M} = 2.431$ eV , $\rho = 3.295$, $r_{0} = 1.763$ \AA.
The fitting procedure performed in Ref. \cite{Lyalin_2009_PhysRevB.79.165403} resulted
in the values of the Ni--C bond dissociation energy $D_e^{\rm Ni-C}$ = 2.43~eV,
equilibrium distance $R_e^{\rm Ni-C} = 1.763$~\AA, and $\beta = 1.869$~\AA$^{-1}$.

In Ref. \cite{Ryu_2010_JPhysChemC.114.2022}, a MD investigation of various noble and
transition-metal clusters interacting with graphite was performed.
It was stated that such an interaction is mainly dominated by a weak van der Waals
force, thus the pairwise Lennard-Jones potential was utilized to model it.
The depth of the potential well in the case of the Ni--C interaction was defined as
$D_e^{\rm Ni-C} = 0.023$~eV.
In Refs. \cite{Shibuta_2007_ComputMaterSci.39.842} and \cite{Lin_2007_JMaterProcessTechnol.192.27},
the values of $D_e^{\rm Ni-C} = 2.478$~eV and 0.1~eV were utilized to model the Ni--C
interaction, respectively.
Thus, a broad range of interatomic parameters for various nickel-carbon systems have
been suggested in the previous studies, although it is hard to conclude from these
studies which parameters should be utilized for the description of the stability and
growth mechanism of nanotubes.
In this paper we elaborate on this issue and justify our choice.

%%%%%%%%%%%%%%%%%%%%%%%%%%%%%%%%%%%%%%%%%%%%%%%%%%%%%%%
\section{Results}
\subsection{Stability and instability of the system}

Following the arguments exploited in Ref. \cite{Lyalin_2009_PhysRevB.79.165403},
in the present study we utilize the pairwise Morse potential to describe the
interaction between nickel and carbon atoms.
In order to investigate the dependence of the system evolution on the parameters
of the potential, the depth of the potential well, $D_e^{\rm Ni-C}$, was varied
in the range from 0.2 to 2.4~eV in steps of 0.2~eV.
The geometrical structure of the system simulated at 400~K is presented in
Figure~\ref{Fig_overview}.

As seen from the figure, stability of the system depends strongly on the binding
energy $D_e^{\rm Ni-C}$ between the nickel and carbon atoms.
Thus, when the Ni--C interaction is covalent but rather weak
($D_e^{\rm Ni-C} = 0.2$~eV, panel (b)), the nanotube placed on top of the nickel
nanoparticle is stable, and the final structure obtained after 5~ns of simulation
resembles the initial structure (cf. panel (a)).
Increasing the interaction energy up to 0.8~eV (panel (c)), the nanotube still
remains stable but its lower part is significantly deformed.
One can also observe the deformation of the nanoparticle structure from the icosahedral
(panel (a)) to the droplet-like one.
Finally, when the dissociation energy of the Ni--C bond exceeds 1.0~eV, the system
becomes unstable, and the nanotube collapses during the first 100~ps of the simulation.
Panel (d) illustrates the structure of the system simulated with the value
$D_e^{\rm Ni-C} = 2.4$~eV, which was obtained by fitting the results of DFT-based
calculations of small NiC$_n$ clusters \cite{Yamaguchi_1999_EPJD.9.385}.

\begin{figure*}
\centering
\resizebox{1.7\columnwidth}{!}{\includegraphics{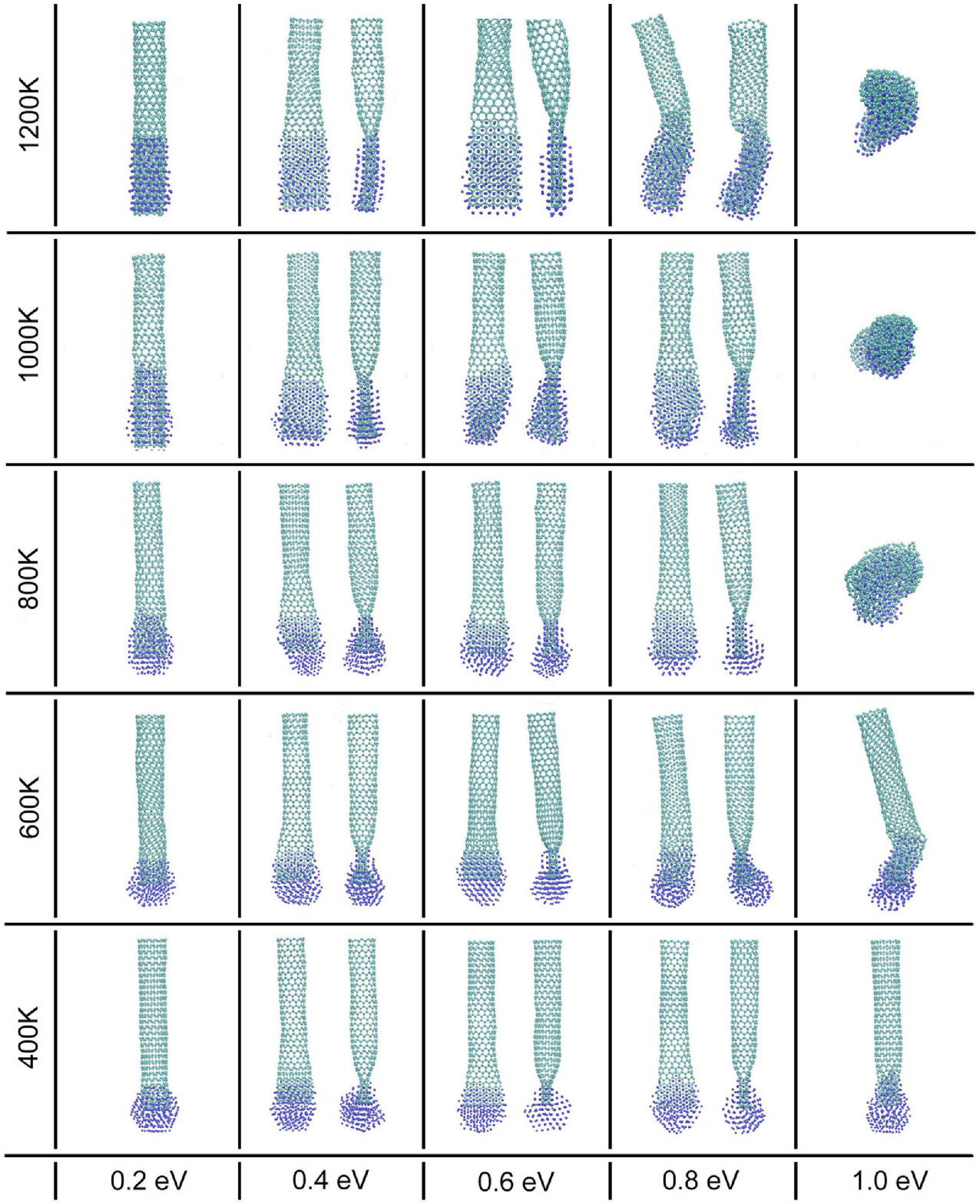}}
\caption{Stability diagram of the single-walled carbon nanotube on top of the nickel
nanoparticle depending on the temperature and the interaction energy between Ni and C atoms.}
\label{Fig_temperatures_all}
\end{figure*}

The results of present simulations are in agreement with the findings obtained on the
basis of the liquid surface model \cite{Solovyov_2008_PhysRevE.78.051601}.
Within this model approach, it was found that if the interaction of a nanotube with a
catalytic nanoparticle is weak (i.e., the well depth of the Ni--C interatomic potential
is less than 1~eV) the longer nanotubes are energetically more favorable.
Since the binding energy per atom decreases in this case, attachment of additional atoms
is energetically favorable resulting in the nanotube growth.
When the nanoparticle-nanotube interaction is strong (e.g., when the well depth
$D_e^{\rm Ni-C} = 1.2$ or 1.5~eV as in Ref.~\cite{Solovyov_2008_PhysRevE.78.051601}),
the trend of the binding energy per atom changes, and it becomes energetically more
favorable for the nanotube to collapse.
This critical value of the Ni--C interaction is in very good agreement to the one,
which we have determined in this work from MD simulations.

As described above, there have been a number of experimental reports during the last decade
on the successful growth of single- and multi-walled carbon nanotubes at various conditions,
in particular, at various temperatures.
In Figure~\ref{Fig_temperatures_all}, we present the results of simulations conducted at
different temperatures in the range from 400 to 1200~K, that corresponds to the typical
range of temperatures at which nanotubes are grown in experiment.

The figure demonstrates that an increase of the Ni--C binding energy leads to the deformation
of the lower part of the nanotube that is in contact with the nanoparticle.
At $T = 600$~K and $D_e^{\rm Ni-C} = 1.0$~eV the nanotube becomes quasistable and bends from
the $z$-axis.
Both the lower part of the nanotube and the nanoparticle lose their initial regular structure
and form a kind of amorphous nickel-carbon system.
Increasing the temperature, the nanotube becomes unstable at $D_e^{\rm Ni-C} = 1.0$~eV.

Therefore, one can conclude that the carbon nanotube remains stable in the whole range
of considered temperatures at comparably large simulation times when the binding energy
between Ni and C atoms is about $0.2 - 0.6$~eV.
However, when the energy is equal to or larger than 0.4~eV, the lower part of the nanotube,
that is in contact with the nanoparticle, is significantly distorted.
Therefore, a highly symmetric structure of the nanotube is kept in the course of simulations
at the Ni--C interaction energies of about 0.2~eV.

%%%%%%%%%%%%%%%%%%%%%%%%%%%%%%%%%%%%%%%%%%%%%%%%%%%%%%%
\subsection{Melting of the catalytic nanoparticle}

\begin{figure}
\centering
\resizebox{1.0\columnwidth}{!}{\includegraphics{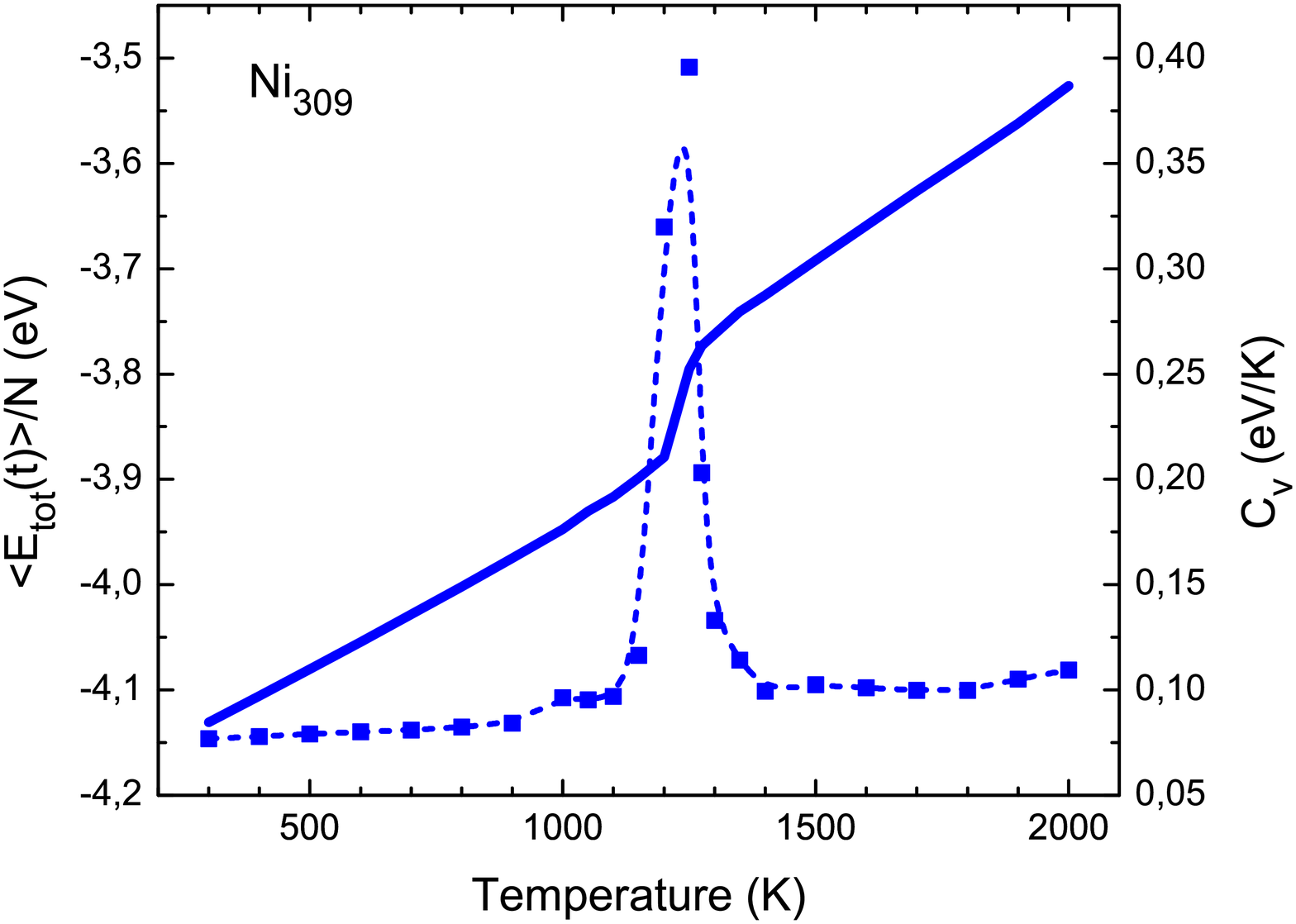}}
\caption{Temperature dependence of the time-averaged total energy $\langle E_{\rm tot} \rangle$
of the isolated Ni$_{309}$ nanoparticle divided by of the number of atoms $N$ (solid curve),
and of the heat capacity $C_V$ (dashed curve and filled squares).
A sharp maximum of the heat capacity at $T = 1240$~K arises at the cluster melting temperature.}
\label{Fig_Ni309_melting}
\end{figure}

When the interaction between nickel and carbon atoms is weak ($D_e^{\rm Ni-C} = 0.2$~eV),
an increase of the temperature leads to the melting of the nanoparticle and its absorption
into the nanotube, as demonstrated in the leftmost column of Figure~\ref{Fig_temperatures_all}.
It is well known that the melting temperature of small metal nanoparticles is significantly
lower as compared to that of bulk materials.
The decrease of the melting temperature of finite-size systems in comparison with the bulk occurs
due to a substantial increase in the relative number of weakly bound atoms on the cluster surface.
According to the so-called Pawlow law \cite{Pawlow_1909_ZPhysChem.65}, the melting temperature $T_m$
of spherical particles possessing a homogeneous surface depends on their radius $R$ as
$T_m = T_m^{\rm bulk}\left( 1 - \alpha/R \right)$, where $T_m^{\rm bulk}$ is the melting temperature
of a bulk material
\cite{Qi_2001_JChemPhys.115.385,Lyalin_2009_PhysRevB.79.165403,Yakubovich_2013_PhysRevB.88.035438}.
%
%Besides the cluster size dependence, the melting temperature should strongly depend on the
%force field, which is used to describe interatomic interaction.
Melting of the isolated Ni$_{309}$ nanoparticle simulated with the Finnis-Sinclair
potential takes place in a broad temperature range between 900~K and 1400~K, as seen from
Figure~\ref{Fig_Ni309_melting}.
There, the caloric curve, i.e. the temperature dependence of the time-averaged total energy
$\langle E_{\rm tot} \rangle$ of the nanoparticle, and the heat capacity at constant volume,
$C_V$, defined as a derivative of the internal energy of the system with respect to temperature,
are shown.

At $T = 1000$ K, the nanoparticle is in the so-called premelted state, when the surface melting
of the cluster takes place, while the cluster core remains in the solid phase and resembles its
crystalline structure.
A small bump in the temperature dependence of the heat capacity indicates for this (see the
dashed curve and filled squares in Figure~\ref{Fig_Ni309_melting}).
%This is a stationary state where the coexistence of two phases, namely liquid surface and a
%frozen core, is observed.
For the Ni$_{309}$ nanoparticle modeled with the Finnis-Sinclair potential the calculated
heat capacity curve has a sharp maximum at $T = 1240$ K, i.e. at the melting temperature of
the cluster.
This value is considerably lower than the melting temperature for the bulk nickel,
$T_m^{\rm bulk} = 1728$K, but fits very well the aforementioned dependence of the melting
temperature of small nanoparticles on their radius.

\begin{figure*}
\centering
\resizebox{2.0\columnwidth}{!}{\includegraphics[scale=0.8,clip]{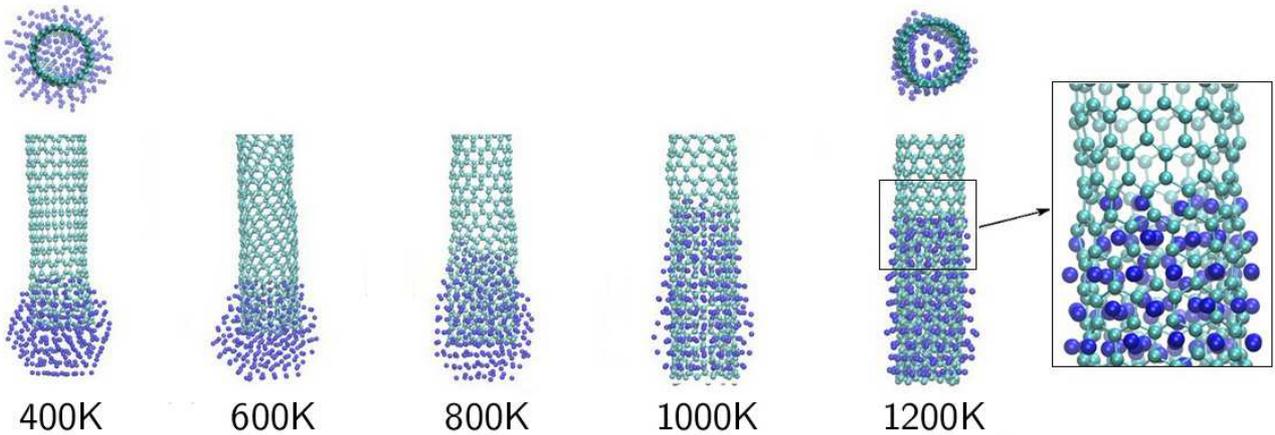}}
\caption{Temperature dependence of the Ni$_{309}$ nanoparticle interacting with a carbon nanotube.
Panels illustrate snapshots of the system after 5~ns of simulation with the binding energy between
Ni and C atoms $D_e^{\rm Ni-C}$ = 0.2~eV.}
\label{Fig_Ni_adsorption}
\end{figure*}

In Ref. \cite{Lyalin_2009_PhysRevB.79.165403,Ding_2004_JVacSciTechnolA.22.1471} it was discussed
that alloying transition-metal nanoparticles with carbon leads to a decrease of their melting
temperature, that can reach several hundred Kelvin at a carbon concentration of about 10\% \cite{Ding_2004_JVacSciTechnolA.22.1471}.
This phenomenon is illustrated in Figure~\ref{Fig_Ni_adsorption}.
At $T=400$ and 600~K, the Ni$_{309}$ nanoparticle remains in the solid phase, while the nanotube
penetrates further into the nanoparticle as the temperature increases.
At $T = 800$ K, there is an evidence of the nanoparticle melting, which initiates its absorption
by the nanotube.
At higher temperatures, $T = 1000$ and 1200~K, the nanoparticle is in the molten state, and nickel
atoms fill the interior of the nanotube.
In this case, the system is stabilized in such a way that nickel atoms are mostly located in the
center of carbon rings, as illustrated in the inset of Figure~\ref{Fig_Ni_adsorption}.

%%%%%%%%%%%%%%%%%%%%%%%%%%%%%%%%%%%%%%%%%%%%%%%%%%%%%%%
\subsection{Validation of parameters of the Ni--C interaction}

\begin{figure*}
\centering
\resizebox{2.0\columnwidth}{!}{\includegraphics{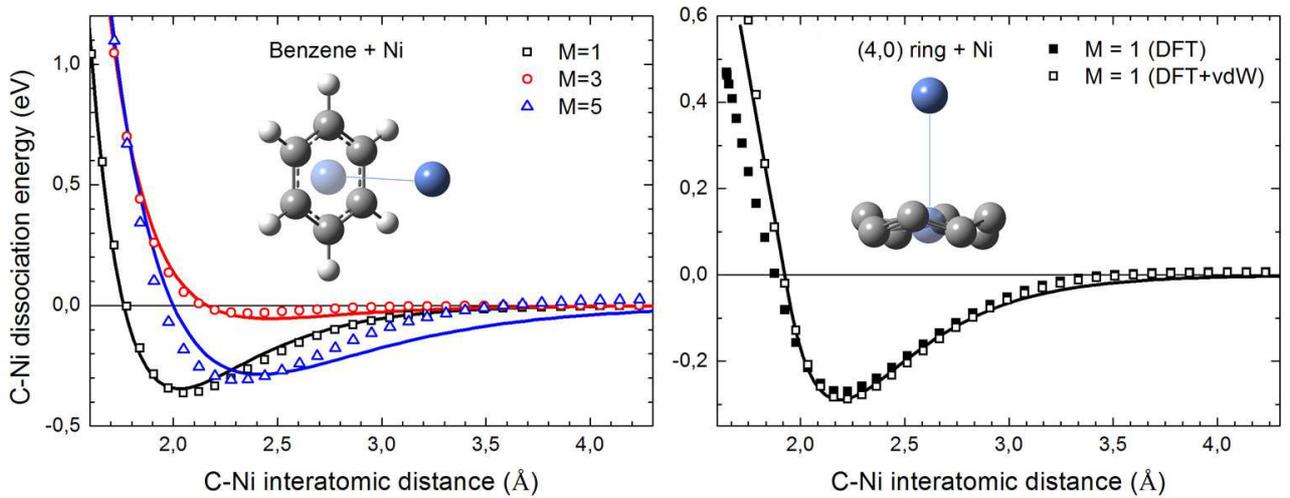}}
\caption{Dependence of the Ni--C bond dissociation energy on the interatomic distance for
the case of the Ni--Benzene (left panel) and Ni--carbon ring (right panel) systems.
Open symbols illustrate the results of DFT calculations performed at B3PW91/AUG-cc-pVTZ
level of theory, corrected by the van der Waals potential.
Solid curves represent fitting of the obtained dependencies with the pairwise Morse potential.
The potential energy curve obtained from the DFT calculations without the van der Waals-type
correction is shown by filled squares.
$M$ stands for multiplicity of the system.}
\label{C-Ni_graph}
\end{figure*}

In order to validate the parameters of interaction energy between Ni and C atoms that
we have utilized in MD simulations, we performed {\it ab initio} calculations of the
potential energy surface for two Ni--C-based compounds.
As a case study, we considered a Ni atom placed in the center of a benzene molecule
and of a small carbon ring, and then moved the metal atom away from the carbon systems
(see Figure~\ref{C-Ni_graph}).
Considering these case studies, we have accounted for different types of Ni--C interaction,
that exist when nickel atoms interact either with the nanotube sidewall or with its open end.

The DFT calculations were performed using the hybrid functional composed of the Becke-type
gradient-corrected exchange and the gradient-corrected correlation functional of Perdew and
Wang (B3PW91) \cite{Becke_1993_JChemPhys.98.5648,Perdew_1992_PhysRevB.46.6671}.
As a basis set, we utilized the correlation-consistent polarized AUG-cc-pVTZ set augmented with
diffuse functions \cite{Dunning_1989_JChemPhys.90.1007,Balabanov_2005_JChemPhys.123.064107}.
The calculations were carried out using Gaussian 09 software package \cite{g09}.

Performing the potential energy scan, we calculated the binding energy of the systems,
\begin{equation}
E_b = E_{\rm Bz/ring} + E_{\rm Ni} - E_{\rm Bz/ring+Ni} \ ,
\end{equation}

\noindent and then divided the obtained value of $E_b$ by the number of carbon atoms.
As a result, we calculated the dissociation energy of a single Ni--C bond.
Such an approach is valid since the benzene molecule and the carbon ring are highly
symmetric structures, and the Ni atom was moved along the main axis of the systems.
Thus, each of the Ni--C bonds may be considered as equivalent.

The results of the potential energy scan performed within the above mentioned DFT framework
for the case of benzene and for a carbon ring are presented in the left and right panels of
Figure~\ref{C-Ni_graph}, respectively.
We considered a small carbon ring consisted of 8 atoms that was cut from a narrow $(4,0)$
nanotube.
The radius of the ring, calculated after geometry optimization, is 1.65~\AA, that is slightly
larger than that of the benzene molecule, $R_{\rm Bz} = 1.40$~\AA.
Since the parameters of the Ni--C interaction may depend on the spin state of the system,
we studied the singlet (multiplicity $M = 1$), triplet ($M=3$) and pentet ($M=5$) states.
In the case of the ring, only the singlet state was analyzed, since the higher spin states
of the system are unstable.

As it was discussed earlier (see Ref. \cite{Obolensky_2007_IntJQuantChem.107.1335} and references
therein), conventional DFT methods cannot account for the van der Waals interaction, which becomes
dominant at large interatomic distances.
In order to get a better description of the van der Waals interaction between Ni and C atoms,
the results of DFT calculations were corrected by a phenomenological van der Waals-type term:
\begin{eqnarray}
E &=& E_{\rm DFT} + E_{\rm vdW}
\nonumber \\
&\equiv& E_{\rm DFT} +
\epsilon \left[ \left( \frac{R_{\rm min}}{r_{ij}} \right)^{12} -
2 \left( \frac{R_{\rm min}}{r_{ij}} \right)^{6} \right] \ .
\end{eqnarray}

\noindent
The $\epsilon \equiv \epsilon({\rm NiC})$ and $R_{\rm min} \equiv R_{\rm min}({\rm NiC})$
functions were constructed as follows:
\begin{eqnarray}
\epsilon({\rm NiC}) &=& \sqrt{\epsilon({\rm Ni}) \cdot \epsilon({\rm C})} \ ,
\nonumber \\
R_{\rm min}({\rm NiC}) &=& \frac{R_{\rm min}({\rm Ni})}{2} + \frac{R_{\rm min}({\rm C})}{2} \ .
\end{eqnarray}

\noindent For the nickel atom, we utilized the values of $\epsilon({\rm Ni})$ = 0.083 eV
and $R_{\rm min}({\rm Ni})$ = 0.63 \AA~\cite{MBN_Explorer1}, while the corresponding
parameters for a carbon atom were taken from the CHARMM22 force field:
$\epsilon({\rm C})$ = 0.00466 eV and $R_{\rm min}({\rm C})$ = 4.0 \AA~\cite{CHARMM22_FF}.

The resulting potential energy curves for the Ni--C interaction, corrected by the
introduced above van der Waals-type term, are marked by open symbols in Figure~\ref{C-Ni_graph}.
Different spin states of the Benzene-Ni compound affect the depth of the potential well
and the equilibrium interatomic distance, ranging from
$D_e$ = 0.36 eV and $R_e$ = 2.05 \AA \ for the singlet state to $D_e$ = 0.05 eV and
$R_e$ = 2.36 \AA \ for the triplet state.
In the case of the carbon ring, the calculated values are $D_e$ = 0.29 eV and $R_e$ = 2.19 \AA.
For the sake of completeness, the potential energy curve obtained from the DFT calculations
without the van der Waals-type correction is shown by filled squares in the right panel
of Figure~\ref{C-Ni_graph}. %for the case of the carbon ring.
This correction does not affect strongly the behavior of the potential energy curve for the
most stable singlet state but should play a much more significant role for the case of less
stable states with higher multiplicities.
For instance, the value of the van der Waals correction at the equilibrium C--Ni distance is
about 0.02~eV which is comparable with $D_e$ = 0.05 eV for the triplet state of the Benzene-Ni
compound.
\begin{table}[h]
\caption{
Parameters of the Morse potential providing the best fit of DFT+vdW potential energy
curves for the Ni--C interaction.}
\begin{center}
%\begin{tabular}{p{3cm}p{2cm}p{2cm}p{2cm}}
\begin{tabular}{p{2.5cm}ccc}
\hline
             &  $D_e^{\rm Ni-C}$ (eV) \qquad  &  $R_e^{\rm Ni-C}$ (\AA) \qquad  &  $\beta$ (\AA$^{-1}$) \\
\hline
Bz+Ni: $M=1$    &          0.345          &             2.03         &   2.597   \\
Bz+Ni: $M=3$    &          0.054          &             2.46         &   2.295   \\
Bz+Ni: $M=5$    &          0.286          &             2.42         &   1.667   \\
ring+Ni: $M=1$  &          0.290          &             2.19         &   2.621   \\
%averaged        &          0.236          &             2.27         &   2.238   \\
\hline
\end{tabular}
\end{center}
\label{tab:tab2}
\end{table}

The calculated potential-energy dependencies were fitted with pairwise Morse potential parametrization;
the resulting curves are shown in Figure~\ref{C-Ni_graph} by solid lines.
The fitting parameters for each case are summarized in Table~\ref{tab:tab2}.
%Averaging over the cases, one derives the mean value of the Ni--C bond dissociation
%energy being equal to 0.236~eV.
%This number
The presented values of the Ni--C bond dissociation energy correspond well to the value
of about 0.2~eV, at which a nanotube atop a nickel nanoparticle turns out to be stable in
MD simulations.
The other parameters utilized in the simulations, namely the equilibrium distance
$R_e^{\rm Ni-C} = 1.763$~\AA~and $\beta = 1.869$~\AA$^{-1}$, were taken from
Ref.~\cite{Lyalin_2009_PhysRevB.79.165403}.
These values are smaller than most of the corresponding numbers obtained from the
DFT calculations (see Table~\ref{tab:tab2}).
However, such a difference in the values of the geometrical characteristics $R_e$
and $\beta$ should not affect significantly the main conclusions made on the basis
of the performed MD simulations, as they are mostly determined by the energetic
characteristics of the system.
This is also supported by the analysis performed in Ref. \cite{Lyalin_2009_PhysRevB.79.165403}
demonstrating
that an increase of the bond length by a factor 1.5, from the equilibrium value up to 2.645~\AA,
does not change the melting temperature of the C-doped Ni$_{147}$ cluster.
Therefore, the difference in 0.5~\AA~between the bond length utilized in MD simulations and
the one obtained from DFT calculations should not affect significantly the conclusions about
the thermodynamic and stability properties of a nanotube placed on the nickel nanoparticle
made in our work on the basis of the performed MD simulations.

%%%%%%%%%%%%%%%%%%%%%%%%%%%%%%%%%%%%%%%%%%%%%%%%%%%%%%%
\section{Conclusion}

In the present study, we have investigated the stability of a carbon nanotube placed on top
of the catalytic Ni$_{309}$ nanoparticle by means of classical molecular dynamics simulations.
An explicit set of constant-temperature simulations has been performed with the Langevin thermostat
in a broad temperature range between 400 and 1200~K, for which a successful growth of single- and
multi-walled nanotubes has been achieved experimentally by means of chemical vapor deposition methods.

In the performed simulations, the nickel-carbon interaction has been modeled by a pairwise
Morse potential.
The influence of the parameters on the thermodynamic and stability properties of the system
has been analyzed.
It was clearly demonstrated that depending on the parameters of the Ni--C potential
different scenarios of the nanotube dynamics atop the nanoparticle are possible.
When the Ni--C interaction is relatively weak and does not exceed several tenths of electronvolt,
the nanotube placed on top of the nanoparticle is stable and resembles its initial structure
in the course of MD simulations.
An increase of the interaction energy causes a significant deformation of both the nanotube
and the nanoparticle even at relatively low temperature, $T=400$~K.
Further increase of the interaction energy leads to the abrupt collapse of the nanotube in
the initial stage of simulation.

By means of MD simulations we have demonstrated that at $T=800$~K the nanoparticle, linked
to the nanotube, premelts; this, in turn, initiates its absorption by the nanotube.
It has been shown that the melting temperature of the isolated Ni$_{309}$ nanoparticle is
several hundred Kelvin higher than it is in the case of the nanoparticle linked to a nanotube.
It is in agreement with previous findings that alloying transition-metal nanoparticles with
carbon leads to a decrease of their melting temperature.

In order to validate the parameters of the Ni--C interaction, utilized in the MD simulations,
we have performed a set of DFT calculations of the potential energy of the Ni--C interaction
for the two Ni--C-based compounds, namely for a Ni atom placed in the center of (i) a benzene
molecule and (ii) a small carbon ring.
The results of DFT calculations have been corrected by accounting for the large-distance
van der Waals interaction between the Ni and C atoms and analyzed for different spin states of
the systems.
The calculated dissociation energy of the Ni--C bond corresponds to the values needed for the
stability and absence of a noticeable deformation of a nanotube atop a nanoparticle at relatively
low temperatures in MD simulations.

In the further work, the validated parameters of the Ni--C interatomic potential can be
utilized for simulating the nanotube formation and growth in the presence of a carbon-rich
feedstock gas.
Usage of the pairwise Morse potential will allow for the study of the atomistic details of
the carbon nanotube growth mechanism, keeping a clear physical picture of the processes
occurring in the system.

%%%%%%%%%%%%%%%%%%%%%%%%%%%%%%%%%%%%%%%%%%%%%%%%%%%%%%%
\section*{Acknowledgement}

The authors acknowledge the Center for Scientific Computing (CSC) of the Goethe University
Frankfurt for the opportunity to carry out complex resource-demanding calculations using
the CPU ''Fuchs'' and CPU/GPU ''LOEWE-CSC'' clusters.

%%%%%%%%%%%%%%%%%%%%%%%%%%%%%%%%%%%%%%%%%%%%%%%%%%%%%%%%%%%%%%%%%%%%%


\begin{thebibliography}{99}

\bibitem{Iijima_1991_Nature.354.56}
   S. Iijima,
   Nature {\bf 354}, 56 (1991)
% [Helical Microtubules of Graphitic Carbon]
% 56--58


\bibitem{Journet_1997_Nature.388.756}
  C. Journet et al.,
% C. Journet, W.K. Maser, P. Bernier, A. Loiseau, M. Lamy de la Chapelle, S. Lefrant,
% P. Deniard, R. Lee, and J.E. Fischer
  Nature {\bf 388}, 756 (1997)
% [Large-scale production of single-walled carbon nanotubes by the electric-arc technique]
% 756--758


\bibitem{Smalley_1996_Science.273.483}
  A. Thess et al.,
% A. Thess, R. Lee, P. Nikolaev, H. Dai, P. Petit, J. Robert, C. Xu, Y.H. Lee, S.G. Kim,
% D.T. Colbert, G. Scuseria, D. Toma´nek, J.E. Fischer, and R.E. Smalley
  Science {\bf 273}, 483 (1996)
% [Crystalline Ropes of Metallic Carbon Nanotubes]
% 483--487


\bibitem{Terrones_1997_Nature.388.52}
  M. Terrones et al.,
% M. Terrones, N. Grobert, J. Olivares, J.P. Zhang, H. Terrones, K. Kordatos, W.K. Hsu,
% J.P. Hare, P.D. Townsend, K. Prassides, A.K. Cheetham, H.W. Kroto, and D.R.M. Walton,
  Nature {\bf 388}, 52 (1997)
% [Controlled production of aligned-nanotube bundles]
% 52--55


\bibitem{Choi_2000_ApplPhysLett.76.2367}
  Y.C. Choi et al.,
% Y.C. Choi, Y.M. Shin, Y.H. Lee, B.S. Leeb, G.-S. Park, W.B. Choi, N.S. Lee, and J.M. Kim
  Appl. Phys. Lett. {\bf 76}, 2367 (2000)
% [Controlling the Diameter, Growth Rate, and Density of Vertically Aligned Carbon Nanotubes
% Synthesized by Microwave Plasma-Enhanced Chemical Vapor Deposition]
% 2367--2369


\bibitem{Li_2008_MaterLett.62.1472}
  H. Li, C. Shi, X. Du, C. He, J. Li, N. Zhao,
  Mater. Lett. {\bf 62}, 1472 (2008)
% [The influence of synthesis temperature and Ni catalyst on the growth of carbon nanotubes
% by chemical vapor deposition]
% 1472--1475


\bibitem{Yuan_2008_NanoLett.8.2576}
  D. Yuan et al.,
% D. Yuan, L. Ding, H. Chu, Y. Feng, T.P. McNicholas, and J. Liu
  Nano Lett. {\bf 8}, 2576 (2008)
% [Horizontally Aligned Single-Walled Carbon Nanotube on Quartz from a Large Variety of Metal
% Catalysts]
% 2576--2579


\bibitem{Kang_2009_JMaterSci.44.2471}
  J. Kang, J. Li, N. Zhao, X. Du, C. Shi, P. Nash,
  J. Mater. Sci. {\bf 44}, 2471 (2009)
% [The effect of catalyst evolution at various temperatures on carbon nanostructures formed
% by chemical vapor deposition]
% 2471--2476


\bibitem{Diao_2010_AdvMater.22.1430}
  P. Diao, Z. Liu,
  Adv. Mater. {\bf 22}, 1430 (2010)
% [Vertically Alligned Single-Walled Carbon Nanotubes by Chemical Assembly - Nethodology,
% Properties, and Applications]
% 1430--1449


\bibitem{Martinez-Limia_2007_JMolModel.13.595}
  A. Martinez-Limia, J. Zhao, P.B. Balbuena,
  J. Mol. Model. {\bf 13}, 595 (2007)
% [Molecular Dynamics Study of the Initial Stages of Catalyzed Single-wall Carbon Nanotubes
% Growth: Force Field Development]
% 595--600


\bibitem{Harris_2007_Carbon.45.229}
  P.J.F. Harris,
  Carbon {\bf 45}, 229 (2007)
% [Solid state growth mechanisms for carbon nanotubes]
% 229--239


\bibitem{Kang_2008_MaterSciEngA.475.136}
  J. Kang, J. Li, X. Du, C. Shi, N. Zhao, P. Nash,
  Mater. Sci. Eng. A {\bf 475}, 136 (2008)
% [Synthesis of carbon nanotubes and carbon onions by CVD using a Ni/Y catalyst supported
% on copper]
% 136--140


\bibitem{SaitoBook}
  R. Saito, G. Dresselhaus, M.S. Dresselhaus,
  {\it Physical Properties of Carbon Nanotubes}
  (Imperial College Press, London, 1998)


\bibitem{Harutyunyan_2005_ApplPhysLett.87.051919}
  A.R. Harutyunyan, T. Tokune, E. Mora,
  Appl. Phys. Lett. {\bf 87}, 051919 (2005)
% [Liquid as a required catalyst phase for carbon single-walled nanotube growth]


\bibitem{Kanzow_1999_PhysRevB.60.11180}
  H. Kanzow, A. Ding,
  Phys. Rev. B {\bf 60}, 11180 (1999)
% [Formation mechanism of single-wall carbon nanotubes on liquid-metal particles]
% 11180--11186


\bibitem{Gavillet_2001_PhysRevLett.87.275504}
  J. Gavillet, A. Loiseau, C. Journet, F. Willaime, F. Ducastelle, J.-C. Charlier,
  Phys. Rev. Lett. {\bf 87}, 275504 (2001)
% [Root-Growth Mechanism for Single-Wall Carbon Nanotubes]


\bibitem{Ding_2004_JChemPhys.121.2775}
  F. Ding, A. Rosen, K. Bolton,
  J. Chem. Phys. {\bf 121}, 2775 (2004)
% [Molecular dynamics study of the catalyst particle size dependence on carbon nanotube growth]
% 2775--2779


\bibitem{Cui_2000_JApplPhys.88.6072}
  H. Cui, O. Zhou, B.R. Stoner,
  J. Appl. Phys. {\bf 88}, 6072 (2000)
% [Deposition of aligned bamboo-like carbon nanotubes via microwave plasma enhanced
% chemical vapor deposition]
% 6072--6074


\bibitem{Jeong_2003_JpnJApplPhys.42.L1340}
  G.-H. Jeong, N. Satake, T. Kato, T. Hirata, R. Hatakeyama, K. Tohji,
  Jpn. J. Appl. Phys. {\bf 42}, L1340 (2003)
% [Time Evolution of Nucleation and Vertical Growth of Carbon Nanotubes during
% Plasma-Enhanced Chemical Vapor Deposition]
% L1340--L1342


\bibitem{Pawlow_1909_ZPhysChem.65}
  P. Pawlow,
  Z. Phys. Chem. {\bf 65}, 1 (1909)
% [\"Uber die Abh\"angigkeit des Schmelzpunktes von der Oberfl\"achenenergie eines festen K\"orpers]
% 1--35


\bibitem{Qi_2001_JChemPhys.115.385}
  Y. Qi, T. \c{C}a\u{g}in, W.L. Johnson, W.A. Goddard III,
  J. Chem. Phys. {\bf 115}, 385 (2001)
% [Melting and Crystallization in Ni Nanoclusters: The Mesoscale Regime]
% 385--394


\bibitem{Lyalin_2009_PhysRevB.79.165403}
  A. Lyalin, A. Hussien, A.V. Solov'yov, W. Greiner,
  Phys. Rev. B {\bf 79}, 165403 (2009)
% [Impurity Effect on the Melting of Nickel Clusters as Seen via Molecular Dynamics Simulations]


\bibitem{Yakubovich_2013_PhysRevB.88.035438}
  A.V. Yakubovich, G.B. Sushko, S. Schramm, A.V. Solov'yov,
  Phys. Rev. B {\bf 88}, 035438 (2013)
% [Kinetics of Liquid-solid Phase Transition in Large Nickel Clusters]


\bibitem{Hofmann_2003_ApplPhysLett.83.135}
  S. Hofmann, C. Ducati, J. Robertson, B. Kleinsorge,
  Appl. Phys. Lett. {\bf 83}, 135 (2003)
% [Low-Temperature Growth of Carbon Nanotubes by Plasma-Enhanced Chemical Vapor Deposition]
% 135--137


\bibitem{Shang_2010_Nanotech.21.505604}
  N.G. Shang, Y.Y. Tan, V. Stolojan, P. Papakonstantinou, S.R.P. Silva,
  Nanotechnology {\bf 21}, 505604 (2010)
% [High-rate low-temperature growth of vertically aligned carbon nanotubes]


\bibitem{He_2011_NanoRes.4.334}
  M. He et al.,
% M. He, A.I. Chernov, E.D. Obraztsova, J. Sainio, E. Rikkinen, H. Jiang, Z. Zhu,
% A. Kaskela, A.G. Nasibulin, E.I. Kauppinen, M. Niemel\"a, and O. Krause
  Nano Res. {\bf 4}, 334 (2011)
% [Low Temperature Growth of SWNTs on a Nickel Catalyst by Thermal Chemical Vapor Deposition]
% 334--342


\bibitem{Halonen_2011_PhysStatSolb.248.2500}
  N. Halonen et al.,
%  N. Halonen, A. S\'{a}pi, L. Nagy, R. Pusk\'{a}s, A.-R. Leino, J. M\"aklin, J. Kukkola,
%  G. T\'{o}th, M.-C. Wu, H.-C. Liao, W.-F. Su, A. Shchukarev, J.-P. Mikkola4, \'{A}. Kukovecz,
%  Z. K\'{o}nya, and K. Kord\'{a}s
  Phys. Stat. Sol. (b) {\bf 248}, 2500 (2011)
% [Low-temperature growth of multi-walled carbon nanotubes by thermal CVD]
% 2500--2503


\bibitem{Irle_2009_NanoRes.2.755}
  S. Irle, Y. Ohta, Y. Okamoto, A.J. Page, Y. Wang, K. Morokuma,
  Nano Res. {\bf 2}, 755 (2009)
% [Milestones in Molecular Dynamic Simulations of Single-Walled Carbon Nanotube Formation:
% A Brief Critical Review]
% 755--767


\bibitem{Ohta_2009_Carbon.47.1270}
  Y. Ohta, Y. Okamoto, S. Irle, K. Morokuma,
  Carbon {\bf 47}, 1270 (2009)
% [Density-functional tight-binding molecular dynamics simulations of SWCNT growth by surface
% carbon diffusion on an iron cluster]
% 1270--1275


\bibitem{Shibuta_2003_ChemPhysLett.382.381}
  Y. Shibuta, S. Maruyama,
  Chem. Phys. Lett. {\bf 382}, 381 (2003)
% [Molecular dynamic simulation of formation process of single-walled carbon nanotubes by CCVD]
% 381--386


\bibitem{Zhao_2005_Nanotechnology.16.S575}
  J. Zhao, A. Martinez-Limia, P.B. Balbuena,
  Nanotechnology {\bf 16}, S575 (2005)
% [Understanding catalysed growth of single-wal carbon nanotubes]
% S575--S581


\bibitem{Tersoff88}
  J. Tersoff,
  Phys. Rev. B {\bf 37}, 6991 (1988)
% [New empirical approach for the structure and energy of covalent systems]
% 6991--7000


\bibitem{Brenner90}
  D.W. Brenner,
  Phys. Rev. B {\bf 42}, 9458 (1990);
  Erratum: {\it ibid.} {\bf 46}, 1948 (1992)
% [Empirical Potential for Hydrocarbons for Use in Simulating the Chemical Vapor Deposition
% of Diamond Films]
% 9458--9471


\bibitem{Smalley_2006_JACS.128.15824}
  R.E. Smalley et al.,
% R.E. Smalley, Y. Li, V.C. Moore, B.K. Price , R. Colorado, Jr., H.K. Schmidt, R.H. Hauge,
% A.R. Barron, and J.M. Tour
  J. Am. Chem. Soc. {\bf 128}, 15824 (2006)
% [Single Wall Carbon Nanotube Amplification: En Route to a Type-Specific Growth Mechanism]
% 15824--15829


\bibitem{Ohta_2008_ACSNano.2.1437}
  Y. Ohta, Y. Okamoto, S. Irle, K. Morokuma,
  ACS Nano {\bf 2}, 1437 (2008)
% [Rapid Growth of a Single-Walled Carbon Nanotube on an Iron Cluster: Density-Functional
% Tight-Binding Molecular Dynamics Simulations]
% 1437--1444


\bibitem{Solovyov_2008_PhysRevE.78.051601}
  I.A. Solov'yov, M. Mathew, A.V. Solov'yov, W. Greiner,
  Phys. Rev. E {\bf 78}, 051601 (2008)
% [Liquid Surface Model for Carbon Nanotube Energetics]


\bibitem{VMD_reference}
  W. Humphrey, A. Dalke, K. Schulten,
  J. Molec. Graphics {\bf 14}, 33 (1996)
% [VMD - Visual Molecular Dynamics]
% 33--38


\bibitem{MBN_Explorer1}
  I.A. Solov'yov, A.V. Yakubovich, P.V. Nikolaev, I. Volkovets, A.V. Solov'yov,
  J. Comput. Chem. {\bf 33}, 2412 (2012)
% [MBN Explorer - A Universal Program for Multiscale Computer Simulations of Complex
%  Molecular Structure and Dynamics]
% 2412--2439

\bibitem{MBN_Explorer2}
  http://www.mbnexplorer.com/


\bibitem{Solovyov_2003_PhysRevLett.90.053401}
  I.A. Solov'yov, A.V. Solov'yov, W. Greiner, A. Koshelev, A. Shutovich,
  Phys. Rev. Lett. {\bf 90}, 053401 (2003)
% [Cluster Growing Process and a Sequence of Magic Numbers]


\bibitem{Geng_2010_PhysRevB.81.214114}
  J. Geng, I.A. Solov'yov, D.G. Reid, P. Skelton, A.E.H. Wheatley, A.V. Solovyov, B.F.G. Johnson,
  Phys. Rev. B {\bf 81}, 214114 (2010)
% [Fullerene-based one-dimensional crystalline nanopolymer formed through topochemical
% transformation of the parent nanowire]


\bibitem{Verkhovtsev_2013_ComputMaterSci.76.20}
  A.V. Verkhovtsev, A.V. Yakubovich, G.B. Sushko, M. Hanauske, A.V. Solov'yov,
  Comput. Mater. Sci. {\bf 76}, 20 (2013)
% [Molecular Dynamics Simulations of the Nanoindentation Process of Titanium Crystal]
% 20--26


\bibitem{Yakubovich_2013_ComputMaterSci.76.60}
  A.V. Yakubovich, A.V. Verkhovtsev, M. Hanauske, A.V. Solov'yov,
  Comput. Mater. Sci. {\bf 76}, 60 (2013)
% [Computer Simulation of Diffusion Process at Interfaces of Nickel and Titanium Crystals]
% 60--64


\bibitem{Sushko_2013_JCompPhys.252.404}
  G.B. Sushko, V.G. Bezchastnov, I.A. Solov'yov, A.V. Korol, W. Greiner, A.V. Solov'yov,
  J. Comput. Phys. {\bf 252}, 404 (2013)
% [Simulation of Ultra-relativistic Electrons and Positrons Channeling in Crystals with MBN Explorer]
% 404--418


\bibitem{Dick_2011_PhysRevB.84.115408}
  V.V. Dick, I.A. Solov'yov, A.V. Solov'yov,
  Phys. Rev. B {\bf 84}, 115408 (2011)
% [Fragmentation Pathways of Nanofractal Structures on Surfaces]


\bibitem{Solovyov_2013_PhysStatSolB}
  I.A. Solov'yov, A.V. Solov'yov, N. K\'{e}baili, A. Masson, C. Br\'{e}chignac,
  Phys. Stat. Sol. (b) {\bf 251}, 609 (2013)
% [Thermally Induced Morphological Transition of Silver Fractals]
% 609--622


\bibitem{Panshenskov_3d-KMC}
  M. Panshenskov, I.A. Solov'yov, A.V. Solov'yov,
  J. Comput. Chem. (2014), DOI: 10.1002/jcc.23613
% [Efficient 3D kinetic monte carlo method for modeling of molecular structure and dynamics]


\bibitem{Finnis-Sinclair}
  M.W. Finnis, J.E. Sinclair,
  Philos. Mag. A {\bf 50}, 45 (1984)
% [A Simple Empirical $N$-body Potential for Transition Metals]
% 45--55


\bibitem{Sushko_2014_JPhysChemA2}
  G.B. Sushko, A.V. Verkhovtsev, A.V. Yakubovich, S. Schramm, A.V. Solov'yov,
  J. Phys. Chem. A (submitted)
% [Molecular Dynamics Simulation of Self-Diffusion Processes in Titanium in Bulk Material,
% on Grain Junctions and on Surface]


\bibitem{Verkhovtsev_2013_ComputTheorChem.1021.101}
  A.V. Verkhovtsev, G.B. Sushko, A.V. Yakubovich, A.V. Solov'yov,
  Comput. Theor. Chem. {\bf 1021}, 101 (2013)
% [Benchmarking of Classical Force Fields by {\it Ab initio} Calculations of Atomic Clusters:
% Ti and Ni-Ti Case]
% 101--108


\bibitem{Sushko_2014_JPhysChemA}
  G.B. Sushko, A.V. Verkhovtsev, A.V. Solov'yov,
  J. Phys. Chem. A (2014), DOI: 10.1021/jp501723w
% [Validation of classical force fields for the description of thermo-mechanical properties
% of transition metal materials]


\bibitem{Gupta}
  R.P. Gupta,
  Phys. Rev. B. {\bf 23}, 6265 (1981)
% [Lattice Relaxation at a Metal Surface]
% 6265--6270


\bibitem{Sutton_Chen}
  A.P. Sutton, J. Chen,
  Philos. Mag. Lett. {\bf 61}, 139 (1990)
% [Long-range Finnis-Sinclair Potentials]
% 139--146


\bibitem{Daw_1993_MaterSciRep.9.251}
  M.S. Daw, S.M. Foiles, M.I. Baskes,
  Mater. Sci. Rep. {\bf 9}, 251 (1993)
% [The embedded-atom method: a review of theory and applications]
%  251--310


\bibitem{Rafii-Tabar_potentials}
  H. Rafii-Tabar, G. A. Mansoori,
%  Interatomic Potential Models for Nanostructures,
  in {\it Encyclopedia of Nanoscience and Nanotechnology, Vol. 4},
  edited by H.S. Nalwa
  (American Scientific Publishers, Valencia, CA, USA, 2004) pp. 231-248


\bibitem{TB-SMA}
  F. Cleri, V. Rosato,
  Phys. Rev. B {\bf 48}, 22 (1993)
% [Tight-binding Potentials for Transition Metals and Alloys]
% 22--33


\bibitem{TB-SMA_2}
  V. Rosato, M. Guellope, B. Legrand,
  Philos. Mag. A {\bf 59}, 321 (1989)
% [Thermodynamical and Structural Properties of FCC Transition-metals Using a Simple Tight-binding Model]
% 321--326


\bibitem{Li_2007_JPhysCondMatter.19.086228}
  J.H. Li, X.D. Dai, T.L. Wang, B.X. Liu,
  J. Phys.: Condens. Matter {\bf 17}, 086228 (2007)
% [A Binomial Truncation Function Proposed for the Second-moment Approximation of Tight-Binding
% Potential and Application in the Ternary Ni–Hf–Ti System]


\bibitem{Daw_1983_PhysRevLett.50.1285}
  M.S. Daw, M.I. Baskes,
  Phys. Rev. Lett. {\bf 50}, 1285 (1983)
% [Semiempirical, Quantum Mechanical Calculation of Hydrogen Embrittlement in Metals]
% 1285--1288


\bibitem{Daw_1984_PhysRevB.29.6443}
  M.S. Daw, M.I. Baskes,
  Phys. Rev. B {\bf 29}, 6443 (1984)
% [Embedded-atom Method: Derivation and Application to Impurities, Surfaces, and Other Defects in Metals]
% 6443--6453


\bibitem{Tomanek_1985_PhysRevB.32.5051}
  D. Tom\'{a}nek, A.A. Aligia, C.A. Balseiro,
  Phys. Rev. B {\bf 32}, 5051 (1985)
% [Calculation of elastic strain and electronic effects on surface segregation]
% 5051--5056


\bibitem{Lai_2000_JPhysCondMatter.12.L53}
  W.S. Lai, B.X. Liu,
  J. Phys.: Condens. Matter {\bf 12}, L53 (2000)
% [Lattice Stability of Some Ni--Ti Alloy Phases Versus Their Chemical Composition and Disordering]
% L53--L60


\bibitem{Yamaguchi_1999_EPJD.9.385}
  Y. Yamaguchi, S. Maruyama,
  Eur. Phys. J. D {\bf 9}, 385 (1999)
% [A Molecular Dynamics Study on the Formation of Metallofullerene]
% 385--388


\bibitem{Shibuta_2007_ComputMaterSci.39.842}
  Y. Shibuta, S. Maruyama,
  Comput. Mater. Sci. {\bf 39}, 842 (2007)
% [Bond-order Potential for Transition Metal Carbide Cluster for the Growth Simulation of a
% Single-walled Carbon Nanotube]
% 842--848


\bibitem{Ryu_2010_JPhysChemC.114.2022}
  J.H. Ryu, H.Y. Kim, D.H. Kim, D.H. Seo, H.M. Lee,
  J. Phys. Chem. C {\bf 114}, 2022 (2010)
% [Immobilization of Au Nanoclusters Supported on Graphite: Molecular Dynamics Simulation]
% 2022--2026


\bibitem{Lin_2007_JMaterProcessTechnol.192.27}
  Z.-C. Lin, J.-C. Huang, Y.-R. Jeng,
  J. Mater. Process. Technol. {\bf 192-193}, 27 (2007)
% [3D nano-scale cutting model for nickel material]
% 27--36


\bibitem{Ding_2004_JVacSciTechnolA.22.1471}
  F. Ding, K. Bolton, A. Ros\'{e}n,
  J. Vac. Sci. Technol. A {\bf 22}, 1471 (2004)
% [Iron-carbide cluster thermal dynamics for catalyzed carbon nanotube growth]


%\bibitem{Perdew_Wang}
%  J.P. Perdew, Y. Wang,
%  Phys. Rev. B {\bf 45}, 13244 (1992)
% [Accurate and Simple Analytic Representation of the Electron-gas Correlation Energy]
% 13244--13249


\bibitem{Becke_1993_JChemPhys.98.5648}
  A.D. Becke,
  J. Chem. Phys. {\bf 98}, 5648 (1993)
% [Density-functional thermochemistry. III. The role of exact exchange]
% 5648--5652


\bibitem{Perdew_1992_PhysRevB.46.6671}
  J.P. Perdew, J.A. Chevary, S.H. Vosko, K.A. Jackson, M.R. Pederson, D.J. Singh, C. Fiolhais,
  Phys. Rev. B {\bf 46}, 6671 (1992)
% [Atoms, Molecules, Solids, and Surfaces: Applications of the Generalized Gradient Approximation
% for Exchange and Correlation]
% 6671--6687


\bibitem{Dunning_1989_JChemPhys.90.1007}
% aug-cc-pVDZ through aug-cc-pVQZ for H, B-Ne
  R.A. Kendall, T.H. Dunning, Jr., R.J. Harrison,
  J. Chem. Phys. {\bf 96}, 6796 (1992)
% [Electron Affinities of the First-Row Atoms Revisited. Systematic Basis Sets and Wave Functions]
% 6796--6806


\bibitem{Balabanov_2005_JChemPhys.123.064107}
% (aug-)cc-p(wC)VnZ-DK, (aug)cc-p(wC)VnZ (n=D-5) for 1st row transition metals
  N.B. Balabanov, K.A. Peterson,
  J. Chem. Phys. {\bf 123}, 064107 (2005)
% [Systematically convergent basis sets for transition metals. I. All-electron correlation
% consistent basis sets for the 3d elements Sc – Zn]


\bibitem{g09}
  M.J. Frisch et al.,
% M.J. Frisch, G.W. Trucks, H.B. Schlegel, G.E. Scuseria, M.A. Robb, J.R. Cheeseman,
% G. Scalmani, V. Barone, B. Mennucci, G.A. Petersson, {\it et al.}
  {\it Gaussian 09}, Revision A.01,
  Gaussian: Wallingford, CT, USA, 2009


\bibitem{Obolensky_2007_IntJQuantChem.107.1335}
  O.I. Obolensky, V.V. Semenikhina, A.V. Solov'yov, W. Greiner,
  Int. J. Quant. Chem. {\bf 107}, 1335 (2007)
% [Interplay of Electrostatic and van der Waals Forces in Coronene Dimer]
% 1335--1343


\bibitem{CHARMM22_FF}
  A.D. MacKerell, Jr. et al.,
% A.D. MacKerell, Jr., D. Bashford, R.L. Dunbrack, Jr., J.D. Evanseck, M.J. Field, S. Fischer,
% J. Gao, H. Guo, S. Ha, D. Joseph-McCarthy, L. Kuchnir, K. Kuczera, F.T.K. Lau, C. Mattos,
% S. Michnick, T. Ngo, D.T. Nguyen, B. Prodhom, W.E. Reiher, III, B. Roux, M. Schlenkrich,
% J.C. Smith, R. Stote, J. Straub, M. Watanabe, J. Wiórkiewicz-Kuczera, D. Yin, and M. Karplus
  J. Phys. Chem. B {\bf 102}, 3586 (1998)
% [All-atom empirical potential for molecular modeling and dynamics studies of proteins]
% 3586–3616

\end{thebibliography}
\end{document}